# Online computation of sparse representations of time varying stimuli using a biologically motivated neural network

**Tao Hu**
*hut@janelia.hhmi.org*

**Dmitri B. Chklovskii**
*mitya@janelia.hhmi.org*

*Howard Hughes Medical Institute, Janelia Farm Research Campus, Ashburn, VA 20147, USA*

**Natural stimuli are highly redundant, possessing significant spatial and temporal correlations. While sparse coding has been proposed as an efficient strategy employed by neural systems to encode sensory stimuli, the underlying mechanisms are still not well understood. Most previous approaches model the neural dynamics by the sparse representation dictionary itself and compute the representation coefficients offline. In reality, faced with the challenge of constantly changing stimuli, neurons must compute the sparse representations dynamically in an online fashion. Here, we describe a leaky linearized Bregman iteration (LLBI) algorithm which computes the time varying sparse representations using a biologically motivated network of leaky rectifying neurons. Compared to previous attempt of dynamic sparse coding, LLBI exploits the temporal correlation of stimuli and demonstrate better performance both in representation error and the smoothness of temporal evolution of sparse coefficients.**

## 1 Introduction

Neural systems face the challenge of constantly changing stimuli due to the volatile natural environment or the self-motion. It was suggested that the brain may employ the strategy of sparse coding [1-3], whereby the sensory input is represented by the strong activation of a relatively small set of neurons. Although both electrophysiological recordings [1] and theoretical arguments [4, 5] demonstrate that most neurons are silent at any given moment [3, 6, 7], how a biologically plausible neural network computes the sparse representations of time varying stimuli and dynamically updates the representation coefficients on the fly remains a mystery.

In neuroscience, earlier approaches model the time varying stimuli, for example the natural image sequences [8-10] or the natural sound [2] as a sparse linear combination of spatial-temporal kernels from a dictionary. The neural dynamics are embedded in the dictionary itself and the sparse representation coefficients are computed offline. Building on the seminal work of Olshausen and Field [7], Rozell et al. have proposed an algorithm for computing sparse representations by a dynamical neural network called Local Competitive Algorithm (LCA) [11]. Although a step towards biological realism, the LCA neglects the temporal correlation of the stimuli and treats the static and time varying input in the same way. Given the importance of tracking and smoothing time varying sparse signals in engineering, it is not surprising that many algorithms have been proposed [12-14]. However, they lack a biologically plausible neural network implementation.

In this paper, we introduce a distributed algorithm called leaky linearized Bregman iteration (LLBI), which computes a time series of sparse redundant representations on the architecture of [11] using leaky rectifying neurons. The LLBI exploits the stimuli temporal correlations and computes the representation coefficients online. By incorporating the Bregman divergence [15] between two consecutive representation coefficients in the objective function to be optimized, LLBI encourages the smooth evolution of representations, which is valuable for higher level neural information processing.

The paper is organized as follows. In §2 we describe the problem of dynamic sparse coding and introduce a probabilistic model of time varying stimuli. In §3, we employ the Follow the Regularized Leader (FTRL) strategy and derive LLBI from the perspective of online learning with discounted loss. We test the numerical performance of LLBI in §4 and compare it to that of soft thresholding LCA (SLCA). Finally, we conclude with the discussion of the advantages of LLBI and future work ( §5).

## 2 Statement of problem

### 2.1 Sparse coding of static stimuli

We start by briefly introducing the sparse coding problem for a statistic input $\boldsymbol{f} \in \mathbb{R}^m$. Given an over- complete dictionary $\boldsymbol{A} \in \mathbb{R}^{m\times n}$ $(n > m)$, a sparse solution $\boldsymbol{u} \in \mathbb{R}^n$ of the equation $\boldsymbol{Au} = \boldsymbol{f}$, can be found by solving the following constrained optimization problem:

$$\min \ \|\boldsymbol{u}\|_1 \ \text{s.t.} \ \boldsymbol{Au} = \boldsymbol{f}, \tag{1}$$

which is known as basis pursuit [16]. In practical applications, where $\boldsymbol{f}$ contains noise, one typically formulates the problem differently, in terms of an unconstrained optimization problem known as the Lasso [17]:

$$\min \frac{1}{2}\|\boldsymbol{Au} - \boldsymbol{f}\|_2^2 + \lambda\|\boldsymbol{u}\|_1, \tag{2}$$

where $\lambda$ is the regularization parameter which controls the trade-off between representation error and sparsity. The choice of regularization by $l_1$ norm assures that the problem both remains convex [18-20] and favors sparse solutions [16, 17].

### 2.2 Dynamic sparse coding of time varying stimuli

Different from the conventional setting with static stimulus $\boldsymbol{f}$, here, we consider the dynamic sparse coding of time varying stimuli $\{\boldsymbol{f}_1, \boldsymbol{f}_2, \dots\dots\}$. At every time step $t$, we are aiming at a sparse representation vector $\boldsymbol{u}_t \coloneqq [u_{1,t}, \dots, u_{n,t}]^T \in \mathbb{R}^n$ which approximates the stimulus,

$$\boldsymbol{f}_t \approx \boldsymbol{Au}_t, \tag{3}$$

where the stimulus $\boldsymbol{f}_t \coloneqq [f_{1,t}, \dots, f_{m,t}]^T \in \mathbb{R}^m$. To find a time series of sparse representations $\{\boldsymbol{u}_1, \boldsymbol{u}_2, \dots\dots\}$, we cannot apply (1) or (2) directly, because the stimuli are constantly changing. There is no time to converge to the steady state solution for each $\boldsymbol{u}_t$. In addition, because of the strong temporal correlation within natural stimuli, it is inefficient to compute each $\boldsymbol{u}_t$ completely from scratch. Therefore, the computation should be performed in an online fashion, with dynamic updating of $\boldsymbol{u}_t$ from old values. Before deriving the proposed online algorithm in Section 3, in order to parameterize the sparse time varying stimuli, we adopt a probabilistic model [12]. We assume the stimuli are slow varying, which means the support sets of $\boldsymbol{f}_t$ or the positions of non-zero coefficients in $\boldsymbol{u}_t$ change slowly over time, and the values of these active coefficients also vary slowly in time. We write the i-th coefficient of $\boldsymbol{u}_t$ as a product of two other hidden variables:

$$u_{i,t} = z_{i,t} x_{i,t}, \tag{4}$$

where $z_{i,t} \in \{0,1\}$ and the vector $\boldsymbol{z}_t \coloneqq [z_{1,t}, \dots z_{n,t}]$ defines the active set of $\boldsymbol{u}$ at time $t$. Since the representation is sparse, most entries of $\boldsymbol{z}_t$ are zeros.

We model the temporal evolution of $\boldsymbol{z}_t$ as a first order Markov process where each coefficient evolves independent of all other coefficients. At each time step, the update is governed by two Markov transition probabilities, $p_{10} \coloneqq \Pr\{z_{i,t} = 1 | z_{i,t-1} = 0\}$ and $p_{01} \coloneqq \Pr\{z_{i,t} = 0 | z_{i,t-1} = 1\}$.

We model the temporal evolution of $\boldsymbol{x}_t \coloneqq [x_{1,t}, \dots x_{n,t}]$ as a first order autoregressive process where the update of each coefficient is independent of all other coefficients and is given by

$$x_{i,t} = \alpha x_{i,t-1} + \sqrt{1-\alpha^2}\varepsilon_{i,t}, \tag{5}$$

where $\varepsilon_{i,t} \in N(0, \sigma^2)$, a normal distribution. The forgetting factor $\alpha \in [0,1]$ controls the temporal correlation length. By construction, the time series $\{x_{i,t}\}$ are exponentially correlated Gaussian random variables with zero mean and variance $\sigma^2$. The correlation time satisfies $\exp(-1/\tau) = \alpha$. At one extreme $\alpha = 1$, $\tau \to \infty$, the coefficient is static, while at the other extreme $\alpha = 0$, $\tau \to 0$, the coefficient is memoryless.

# 3 Derivation of LLBI and neural network implementation

## 3.1 Online learning framework and FTRL

We derive LLBI under the framework of online learning. A typical online learning problem is defined as follows [21, 22]:

At each time step $t$ = 1…T

Algorithm make a prediction, $\boldsymbol{u}_t \in \mathbb{R}^n$.

Adversary picks loss function, $\ell_t$,

Algorithm suffers loss $\ell_t(\boldsymbol{u}_t)$ and observes function $\ell_t$ .

The goal is to minimize regret which is defined as:

$$R_T = \sum_{t=1}^{T} \ell_t(\boldsymbol{u}_t) - \min_{\boldsymbol{u}\in\mathbb{R}^n} \sum_{t=1}^{T} \ell_t(\boldsymbol{u}), \tag{6}$$

where $\boldsymbol{u}$ is minimized in hindsight with the knowledge of $\ell_t$ for $t$ = 1…T. We assume that the loss, $\ell_t$, is convex and differentiable allowing us to use convex optimization methods.

A common strategy to minimize regret is the so called Follow the Regularized Leader (FTRL):

$$\boldsymbol{u}_{t+1} = \operatorname{argmin}_{\boldsymbol{u}}\{\eta \sum_{s=1}^{t} \ell_s(\boldsymbol{f}_s, \boldsymbol{u}) + \Psi(\boldsymbol{u})\}, \tag{7}$$

where $\eta > 0$ is the learning rate and $\Psi$ is a convex regularizer. Furthermore, the first prediction is $\boldsymbol{u}_1 = \operatorname{argmin}_{\boldsymbol{u}}\{\Psi(\boldsymbol{u})\}$. Therefore, $\Psi(\boldsymbol{u})$ can be thought to represent our knowledge about $\boldsymbol{u}$ prior to the first time step.

Recently, much attention has been devoted to sparsity inducing regularizers [16, 17, 23], such as the $l_1$ norm, $\Psi(\boldsymbol{u}) = \|\boldsymbol{u}\|_1$. Because the $l_1$ norm is not differentiable at zero, conventional stochastic gradient descent method failed to produce sparse solutions. Several algorithms, such as FOBOS[24], RDA [25], COMID [26], TG [27], have been developed to handle the $l_1$ regularized online learning problem, and many of them can be interpreted in the framework of FTRL [28].

## 3.2 Derivation of LLBI

Below, we derive LLBI from the FTRL framework and implement it using a biologically motivated neural network of leaky rectifying neurons (Fig. 1b). We use a loss function $\ell_s(\boldsymbol{f}_s, \boldsymbol{u}) = \frac{1}{2}\|\boldsymbol{f}_s - \boldsymbol{A}\boldsymbol{u}\|_2^2$, which is the square representation error for the stimulus $\boldsymbol{f}_s$, and an $l_1 - l_2$ norm regularizer, $\Psi(\boldsymbol{u}) = \gamma\|\boldsymbol{u}\|_1 + \frac{1}{2}\|\boldsymbol{u}\|_2^2$ , known as the elastic net [23]. Since the time varying stimuli are exponentially correlated with forgetting factor $\alpha$, recent stimulus is more informative about the updating of $\boldsymbol{u}$. Therefore we apply FTRL with exponentially discounted loss and obtain:

$$\boldsymbol{u}_{t+1} = \operatorname{argmin}_{\boldsymbol{u}}\{\eta \sum_{s=1}^{t} \alpha^{t-s} \ell_s(\boldsymbol{f}_s, \boldsymbol{u}) + \Psi(\boldsymbol{u})\}. \tag{8}$$

The convex cost function $\ell_s$ is usually approximated by a linear form of its gradients [22], $g_s(\boldsymbol{u}_s) = \nabla l_s(\boldsymbol{u}_s)$ and (8) becomes

$$\boldsymbol{u}_{t+1} = \operatorname{argmin}_{\boldsymbol{u}}\{\eta \sum_{s=1}^{t} \alpha^{t-s}\langle g_s(\boldsymbol{u}_s), \boldsymbol{u} - \boldsymbol{u}_s\rangle + \gamma\|\boldsymbol{u}\|_1 + \tfrac{1}{2}\|\boldsymbol{u}\|_2^2\}, \tag{9}$$

where

$$g_s(\boldsymbol{u}_s) = \nabla l_s(\boldsymbol{u}_s) = -\boldsymbol{A}^T(\boldsymbol{f}_s - \boldsymbol{A}\boldsymbol{u}_s). \tag{10}$$

The optimality condition for (9) is:

$$\boldsymbol{v}_{t+1} \in \partial\left[\gamma\|\boldsymbol{u}_{t+1}\|_1 + \tfrac{1}{2}\|\boldsymbol{u}_{t+1}\|_2^2\right], \tag{11}$$

where $\partial[.]$ designates a sub-differential and $-\boldsymbol{v}_{t+1}$ is the gradient of the first term in (9):

$$\boldsymbol{v}_{t+1} = -\eta \textstyle\sum_{s=1}^{t} \alpha^{t-s} g_s(\boldsymbol{u}_s). \tag{12}$$

Similarly, from the condition of optimality for $\boldsymbol{u}_t$

$$\boldsymbol{v}_t = -\eta \textstyle\sum_{s=1}^{t-1} \alpha^{t-1-s} g_s(\boldsymbol{u}_s). \tag{13}$$

By combining (12) and (13), we get

$$\boldsymbol{v}_{t+1} = \alpha\boldsymbol{v}_t - \eta g_t(\boldsymbol{u}_t). \tag{14}$$

Substituting (12) into (9) and simplifying, we obtain

$$\boldsymbol{u}_{t+1} = \mathrm{argmin}_{\boldsymbol{u}}\left\{-\langle\boldsymbol{v}_{t+1},\boldsymbol{u}\rangle + \gamma\|\boldsymbol{u}\|_1 + \tfrac{1}{2}\|\boldsymbol{u}\|_2^2\right\} = \mathrm{argmin}_{\boldsymbol{u}}\left\{\tfrac{1}{2}\|\boldsymbol{u}-\boldsymbol{v}_{t+1}\|_2^2 + \gamma\|\boldsymbol{u}\|_1\right\}. \tag{15}$$

Such minimization problem can be solved using component-wise shrinkage (soft-thresholding) operation [29, 30],

$$u_{i,t} = \mathrm{shrink}(v_{i,t},\gamma) = \begin{cases} v_{i,t}-\gamma, & \text{if } v_{i,t} > \gamma \\ 0, & \text{if } -\gamma \le v_{i,t} \le \gamma, \\ v_{i,t}+\gamma, & \text{if } v_{i,t} < -\gamma \end{cases} \tag{16}$$

which is shown in Fig. 1a. By combining (10), (14) and (15), we arrive at the following:

**Algorithm** Leaky linearized Bregman iteration (LLBI)

**initialize**: $\boldsymbol{v}_1$=0 and $\boldsymbol{u}_1$=0.
**for** $t$ = 1,2,3,…**do**

$$\boldsymbol{v}_{t+1} = \alpha\boldsymbol{v}_t + \eta\boldsymbol{A}^T(\boldsymbol{f}_t - \boldsymbol{A}\boldsymbol{u}_t), \tag{17}$$

$$\boldsymbol{u}_{t+1} = \mathrm{shrink}(\boldsymbol{v}_{t+1},\gamma), \tag{18}$$

**end for**

The reason for the algorithm name becomes clear if we substitute (12) and (14) into (9) and obtain:

$$\boldsymbol{u}_{t+1} = \mathrm{argmin}_{\boldsymbol{u}}\{\eta\langle g_t(\boldsymbol{u}_t),\boldsymbol{u}-\boldsymbol{u}_t\rangle + \gamma\|\boldsymbol{u}\|_1 + \tfrac{1}{2}\|\boldsymbol{u}\|_2^2 - \alpha\langle\boldsymbol{v}_t,\boldsymbol{u}-\boldsymbol{u}_t\rangle\}, \tag{19}$$

which can be written as:

$$\boldsymbol{u}_{t+1} = \mathrm{argmin}_{\boldsymbol{u}}\left\{\eta\langle g_t(\boldsymbol{u}_t),\boldsymbol{u}-\boldsymbol{u}_t\rangle + \alpha D_{\Psi}^{\boldsymbol{v}_t}(\boldsymbol{u},\boldsymbol{u}_t) + (1-\alpha)\Psi(\boldsymbol{u})\right\}, \tag{20}$$

where

$$D_{\Psi}^{\boldsymbol{v}_t}(\boldsymbol{u},\boldsymbol{u}_t) = \Psi(\boldsymbol{u}) - \Psi(\boldsymbol{u}_t) - \langle\boldsymbol{v}_t,\boldsymbol{u}-\boldsymbol{u}_t\rangle \tag{21}$$

is the Bregman divergence induced by the convex function $\Psi(\boldsymbol{u})$ at point $\boldsymbol{u}_t$ for sub-gradient $\boldsymbol{v}_t$ [15]. In fact, (20) can be a starting point for the derivation of LLBI, thus justifying the name. When $\alpha = 1$, (14) and (20) go back to the linearized Bregman iteration originally proposed by Osher and co-authors [31-33] for solving deterministic convex optimization problems, such as basis pursuit [16]. Since the Bregman divergence is convex with respect to $\boldsymbol{u}$, $D_{\Psi}^{\boldsymbol{v}_t}(\boldsymbol{u},\boldsymbol{u}_t) \ge 0$ and equals 0 if and only if $\boldsymbol{u} = \boldsymbol{u}_t$, minimizing also the Bregman divergence in (20) forces $\boldsymbol{u}_{t+1}$ to be close to $\boldsymbol{u}_t$, which results in a smooth evolution of sparse representation coefficients. The forgetting factor $\alpha$ adjusts the attraction strength towards old values of $\boldsymbol{u}$, and should be tuned to match the changing speed of stimuli. We demonstrated this attraction effect of LLBI numerically in Sec. 4.

### 3.3 Biologically motivated neural network implementation of LLBI

The LLBI can be naturally implemented by a neural network of *n* parallel leaky rectifying neurons, Fig. 1b, an architecture previously proposed to implement SLCA [11]. Such a network combines feedforward projections, $\boldsymbol{A}^T$, and inhibitory lateral connections, $-\boldsymbol{A}^T\boldsymbol{A}$, which implement "explaining away" [34]. At every step, each neuron updates its component of the internal variable, $\boldsymbol{v}$ (sub-threshold membrane voltage), by adding the corresponding components of the feedforward signal, $\boldsymbol{A}^T\boldsymbol{f}$, and the broadcast external variable, $-\boldsymbol{A}^T\boldsymbol{A}\boldsymbol{u}$. Then, each neuron computes the new value of its component in $\boldsymbol{u}$ (firing rate) by shrinking its component in $\boldsymbol{v}$. The forgetting factor $\alpha < 1$ naturally maps to the finite membrane time constant $\tau_m$ of a leaky neuron via $\alpha = \exp(-1/\tau_m)$.

Unlike thresholding in the LLBI nodes (16), in biological neurons, thresholding is one-sided [35, 36]. Such discrepancy is easily resolved by substituting each node with two opposing (on- and off-) nodes. In fact, neurons in some brain areas are known to come in two types (on- and off-) [37].

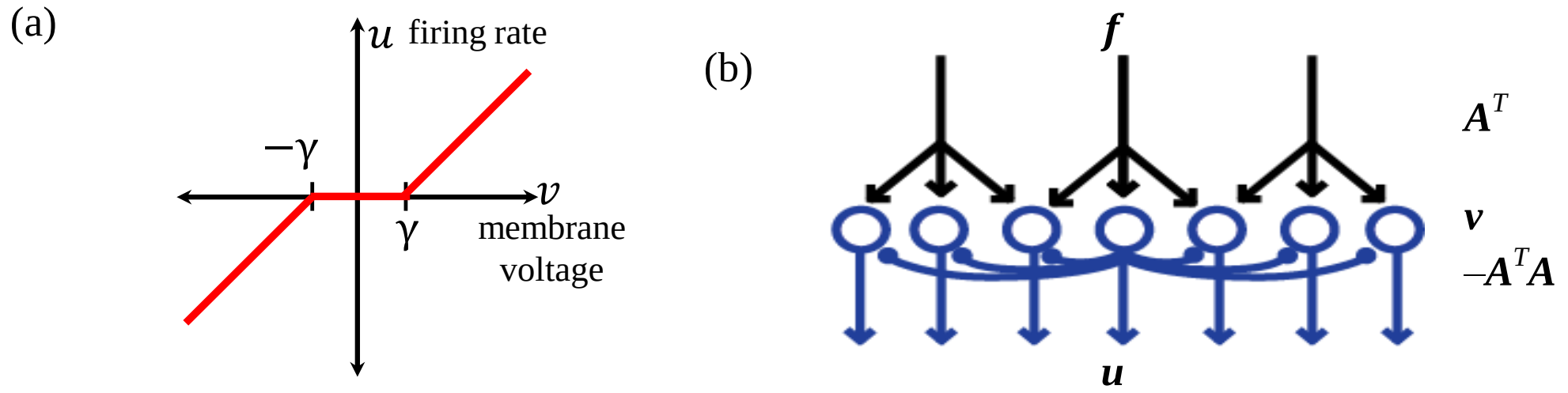

Figure 1: (a) Shrinkage function; (b) A network architecture for SLCA or LLBI. Feedforward projections multiply the input $\boldsymbol{f}$ by a matrix $\boldsymbol{A}^T$, while lateral connections update internal neural activity $\boldsymbol{v}$ (membrane voltage) by a product of matrix $-\boldsymbol{A}^T\boldsymbol{A}$ and external activity $\boldsymbol{u}$ (firing rate).

# 4 Numerical experiments

In this section, we report the numerical performance of LLBI and compare it to that of SLCA [11] which also has a neural network implementation (Fig. 1b).

## 4.1 Experiment on sparse time varying stimuli with invariant support sets

We compute the time series of sparse representations $\{\boldsymbol{u}_t\}$ of synthesized stimuli as described in Sec. 2. By setting $p_{01} = p_{10} = 0$, we first consider the stimuli with time invariant support sets. The elements of the matrix $\boldsymbol{A} \in \mathbb{R}^{64\times128}$ are chosen i.i.d. from a normal distribution $N(0,1)$ and column-normalized by dividing each element by the $l_2$ norm of its column. We construct the time series $\{\bar{\boldsymbol{f}}_t\}$ as $\{\boldsymbol{A}\bar{\boldsymbol{u}}_t\}$, where the initial value $\bar{\boldsymbol{u}}_0 \in \mathbb{R}^{128}$ is generated by randomly selecting $nz = 10$ locations for non-zero entries sampled i.i.d. from a normal distribution $N(0,1)$. We model the temporal evolution of the values of non-zeros entries with forgetting factor $\alpha = 0.99$, variance $\sigma^2 = 1$. Then, we apply LLBI using the network (Fig. 1b) with 128 nodes. For simplicity, we set the learning rate $\eta$=0.99, which is the same as $\alpha$. Then we empirically set the firing threshold $\lambda = 3.1$ to make the mean sparsity of the computed sparse representation $\langle \|\boldsymbol{u}_t\|_0 \rangle$ close to the true sparsity $nz = 10$. For SLCA, we choose the parameters which yield the same mean sparsity and the lowest relative mean square error (MSE) $\langle E_{\boldsymbol{f}_t} \rangle = \langle \|\bar{\boldsymbol{f}}_t - \boldsymbol{f}_t\|_2^2 / \|\bar{\boldsymbol{f}}_t\|_2^2 \rangle$. The reconstructed stimulus is given by $\boldsymbol{f}_t = \boldsymbol{A}\boldsymbol{u}_t$. We also compute the relative MSE for the sparse representation coefficients $\langle E_{\boldsymbol{u}_t} \rangle = \langle \|\bar{\boldsymbol{u}}_t - \boldsymbol{u}_t\|_2^2 / \|\bar{\boldsymbol{u}}_t\|_2^2 \rangle$, The results are shown in Fig. 2. Compared to SLCA, LLBI achieves lower relative MSE both for the recovered stimulus and the recovered representation coefficients. The performance gain of LLBI is obtained by exploiting the temporal correlation of the stimuli and matching the forgetting factor or equivalently the neural membrane time constant to the stimuli characteristic time scale.

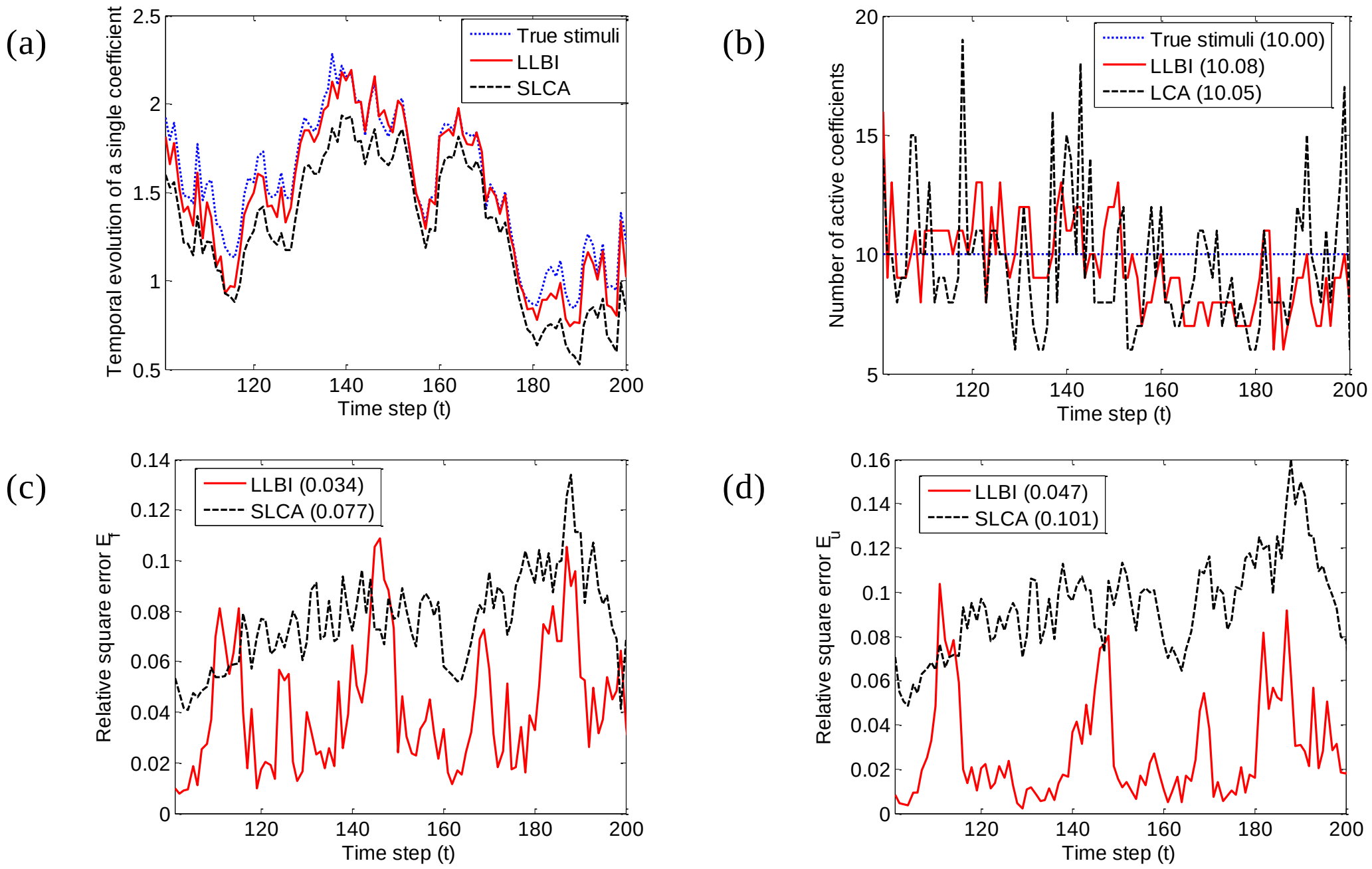

Figure 2: Computing sparse representations of exponentially correlated time varying stimuli with time invariant support sets using LLBI and SLCA. Temporal evolutions of a single active representation coefficient $u_{i,t}$ (a), number of active representation coefficients (b), relative square error $E_{\boldsymbol{f_t}} = \|\bar{\boldsymbol{f}}_t - \boldsymbol{f}_t\|_2^2 / \|\bar{\boldsymbol{f}}_t\|_2^2$ (c) and $E_{\boldsymbol{u_t}} = \|\bar{\boldsymbol{u}}_t - \boldsymbol{u}_t\|_2^2 / \|\bar{\boldsymbol{u}}_t\|_2^2$ (d). The means are shown in the figure legends.

## 4.2 Experiment on sparse time varying stimuli with slow varying support sets

The settings for the second experiment are the same as the first one except that now we allow slow variation of the support sets over time. We set $p_{01} = 0.01$ and $p_{01} = p_{01} \times \frac{nz}{n-nz} \approx 0.008$. With this setting, the transition from active representation coefficients to inactive ones is balanced by the reverse transition, protecting the number of active coefficients from drifting over time. As demonstrated in Fig. 3, LLBI shows better performance against SLCA.

## 4.3 Experiment on the "foreman" test video

We apply LLBI to the first 100 frames of the standard "foreman" test video [38] and compare its performance to that of SLCA. We crop and rescale each video frame to $32 \times 32$ pixels, and normalize the video to have zero mean and unit norm. Then we construct a four time over-complete DCT matrix $\boldsymbol{A} \in \mathbb{R}^{1024\times 4096}$. To compare the performance of LLBI to SLCA, we follow the settings in [11], where each frame is presented to LLBI or SLCA for 33 iterations before switching to the next frame in the video sequence. For SLCA, we use the parameters given in [11] and find the mean number of active representation coefficients for each frame is about 510 (Fig. 4a). We choose the LLBI parameters to achieve similar representation sparsity (Fig. 4a). Specifically, we set $\alpha = \exp(-1/600) \approx 0.998$, because the temporal correlation length of the video is about 20 frames (Fig. 5), which gives a correlation time of about 600 frames when each frame is presented for 33 iterations. We set the other two LLBI parameters as follows, learning rate $\eta = 0.13$ and threshold $\lambda = 0.2$. We observe that LLBI gives lower representation error than SLCA (Fig. 4b). We also notice that the evolution of sparse representation coefficients obtained from LLBI is smoother than that from SLCA, as demonstrated by the smaller number of changed locations of active coefficients (Fig. 4c), which are defined as the total number of neurons changing from active to silent or from silent to active. The improvement of smoothness can also be seen from the smaller ratio between the number of changed locations and active coefficients (Fig. 4d). As explained in Sec. 3.2, the smooth evolution is due to the effective attraction induced by minimizing the Bregman divergence between two consecutive sparse representation coefficients.

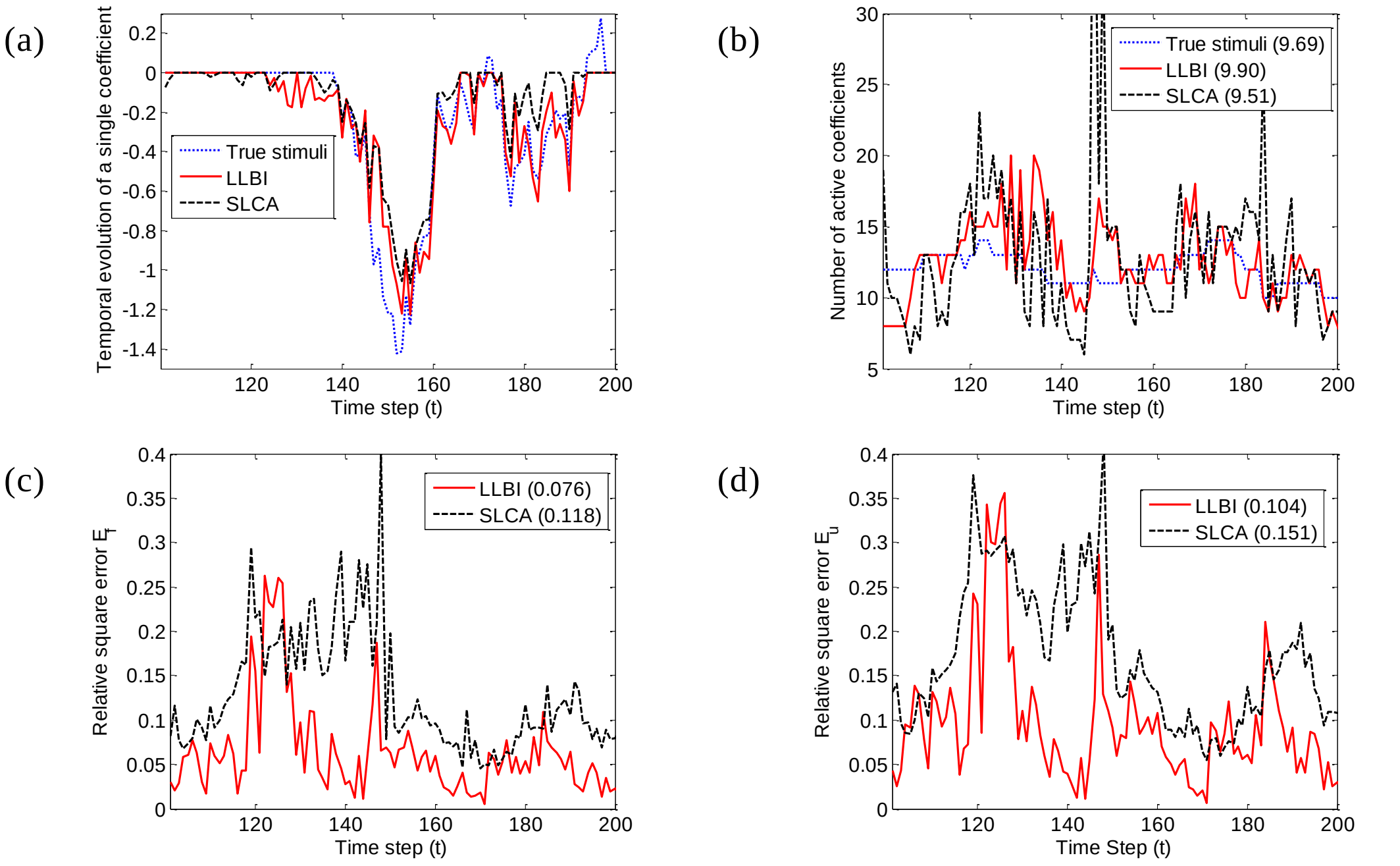

Figure 3: Computing sparse representations of exponentially correlated time varying stimuli with slow varying support sets using LLBI and SLCA. Temporal evolutions of a single active representation coefficient $u_{i,t}$ (a), number of active representation coefficients (b), relative square error $E_{\boldsymbol{f_t}} = \|\bar{\boldsymbol{f}}_t - \boldsymbol{f}_t\|_2^2 / \|\bar{\boldsymbol{f}}_t\|_2^2$ (c) and $E_{\boldsymbol{u_t}} = \|\bar{\boldsymbol{u}}_t - \boldsymbol{u}_t\|_2^2 / \|\bar{\boldsymbol{u}}_t\|_2^2$ (d). The means are shown in the figure legends.

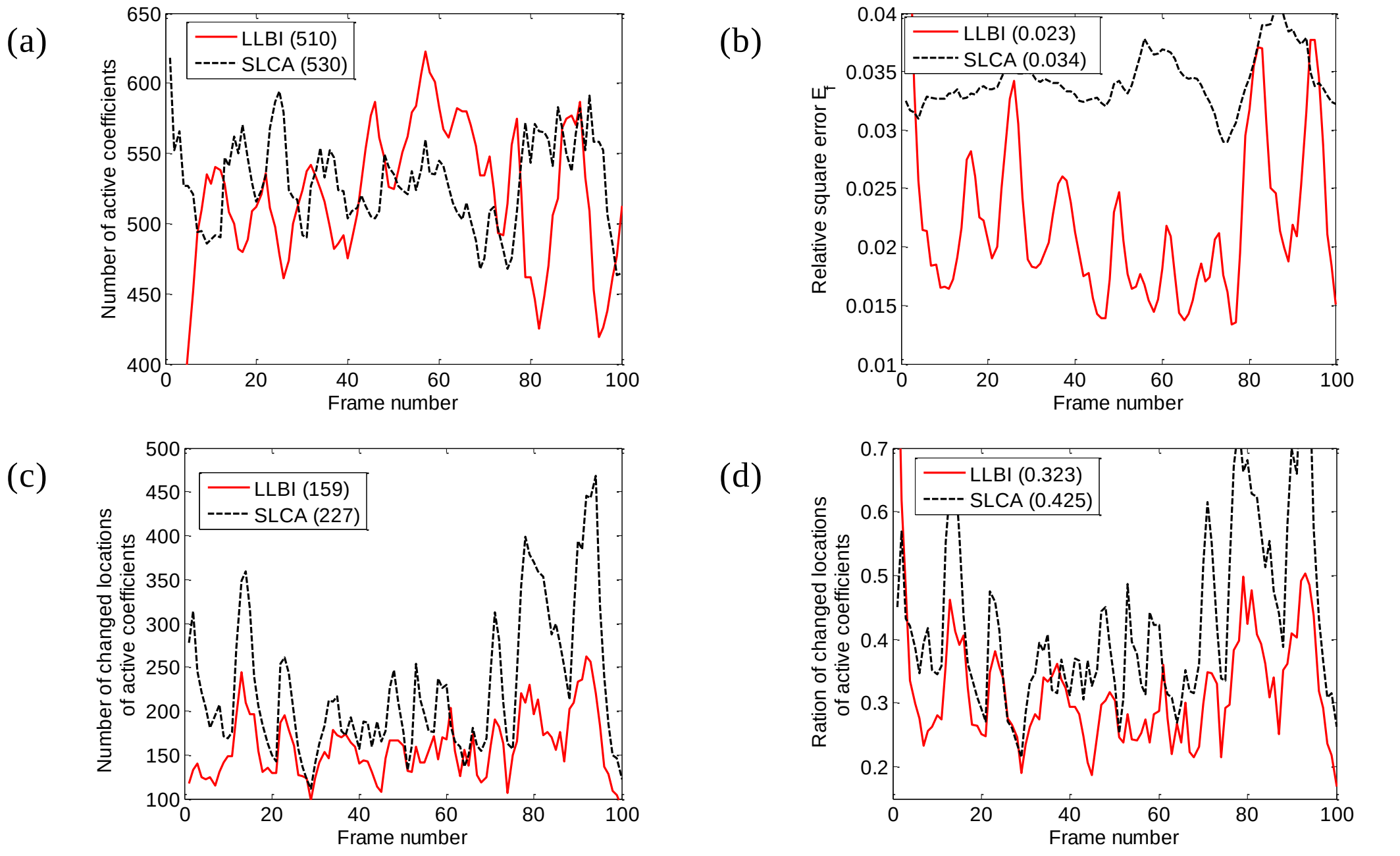

Figure 4: Computing sparse representations of the first 100 frames of the “foreman” test video using LLBI and SLCA. Temporal evolutions of the number of active representation coefficients (a), relative square error $E_{\boldsymbol{f_t}} = \|\bar{\boldsymbol{f}}_t - \boldsymbol{f}_t\|_2^2 / \|\bar{\boldsymbol{f}}_t\|_2^2$ (b), the number of changed locations of active representation coefficients between two consecutive frames (c) and the ratio between the number of changed locations and active coefficients (d). The means are shown in the figure legends.

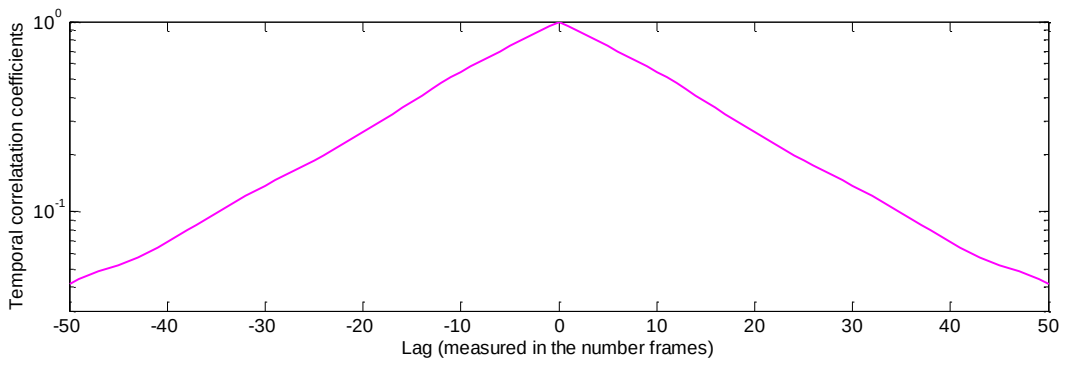

Figure 5: The temporal correlation of “foreman” test video is exponential.

# 5 Summary

In this paper, we propose an algorithm called LLBI, which computes sparse representations of time varying stimuli using a biologically motivated neural network of leaky rectifying neurons. The network implements LLBI in an online fashion with dynamic update of sparse representation coefficients from old values when exposed to the stream of stimuli. Compared to the existing distributed algorithm SLCA which has a neural network implementation (Fig. 1b), LLBI exploits the temporal correlation of time varying signal and thus enjoys smaller representation error and smoother temporal evolution of sparse representations, which are valuable for higher order neural interpretation and processing of sensory stimuli.

Here, we consider only exponentially correlated stimuli possessing only one time scale and require the membrane time constant of neurons matched to the stimuli correlation time. In the future work, we would like to extend the framework of LLBI to the more complicated case of natural stimuli with temporal correlations over multiple time scales. For this case, a network consists of neurons with different membrane time constants (multiple forgetting factors) are highly plausible.

Finally, the LLBI is not only served a model for neural computation, but also a highly promising algorithm for distributed hardware implementations for energy constrained applications which favor online computation of sparse representations.

## Acknowledgments

## References

[1] M. R. DeWeese, *et al.*, "Binary spiking in auditory cortex," *Journal of Neuroscience,* vol. 23, pp. 7940-7949, Aug 27 2003.

[2] M. S. Lewicki, "Efficient coding of natural sounds," *Nat Neurosci,* vol. 5, pp. 356-63, Apr 2002.

[3] B. A. Olshausen and D. J. Field, "Sparse coding of sensory inputs," *Curr Opin Neurobiol,* vol. 14, pp. 481-7, Aug 2004.

[4] D. Attwell and S. B. Laughlin, "An energy budget for signaling in the grey matter of the brain," *J Cereb Blood Flow Metab,* vol. 21, pp. 1133-45, Oct 2001.

[5] P. Lennie, "The cost of cortical computation," *Current Biology,* vol. 13, pp. 493-7, Mar 18 2003.

[6] J. L. Gallant and W. E. Vinje, "Sparse coding and decorrelation in primary visual cortex during natural vision," *Science,* vol. 287, pp. 1273-1276, Feb 18 2000.

[7] B. A. Olshausen and D. J. Field, "Emergence of simple-cell receptive field properties by learning a sparse code for natural images," *Nature,* vol. 381, pp. 607-9, Jun 13 1996.

[8] J. H. van Hateren and D. L. Ruderman, "Independent component analysis of natural image sequences yields spatio-temporal filters similar to simple cells in primary visual cortex," *Proc Biol Sci,* vol. 265, pp. 2315-20, Dec 7 1998.

[9] M. S. Lewicki and T. J. Sejnowski, "Coding time-varying signals using sparse, shift-invariant representations," in *NIPS 11*, 1999.

[10] B. A. Olshausen, "Learning sparse, overcomplete representations of time-varying natural images," in *IEEE International Conference on Image Processing*, 2003.

[11] C. J. Rozell, *et al.*, "Sparse coding via thresholding and local competition in neural circuits," *Neural Comput,* vol. 20, pp. 2526-63, Oct 2008.

[12] J. Ziniel, *et al.*, "Tracking and smoothing of Time-Varying Sparse Signals via Approximate Belief Propagation," in *Proc. of the 44th Asilomar Conference on Signals, Systems and Computers*, 2010, pp. 808-812.

[13] D. Angelosante, *et al.*, "Lasso-Kalman smoother for tracking sparse signals," presented at the Asilomar Conf. Signals, Syst., & Comput., 2009.

[14] D. Angelosante and G. Giannakis, "RLS-weighted Lasso for adaptive estimation of sparse signals," presented at the ICASSP, 2009.

[15] L. M. Bregman, "The relaxation method of finding the common point of convex sets and its application to the solution of problems in convex programming," *USSR Computational Mathematics and Mathematical Physics,* vol. 7, pp. 200-217, 1967.

[16] S. S. B. Chen, *et al.*, "Atomic decomposition by basis pursuit," *Siam Journal on Scientific Computing,* vol. 20, pp. 33-61, 1998.

[17] R. Tibshirani, "Regression shrinkage and selection via the Lasso," *Journal of the Royal Statistical Society Series B-Methodological,* vol. 58, pp. 267-288, 1996.

[18] J. Dattorro, *Convex Optimization & Euclidean Distance Geometry*. Palo Alto, CA: Meboo Publishing, 2008.

[19] D. P. Bertsekas, *Convex optimization theory*. Belmont, MA: Athena Scientific, 2009.

[20] S. Boyd and L. Vandenberghe, *Convex Optimization*. Cambridge, U.K.: Cambridge Univ. Press, 2004.

[21] N. Cesa-Bianchi and G. Lugosi, *Prediction, Learning, and Games*: Cambridge University Press, 2006.

[22] A. Rakhlin. (2009). *Lecture Notes on Online Learning*. Available: http://www-stat.wharton.upenn.edu/~rakhlin/courses/stat991/papers/lecture_notes.pdf

[23] H. Zou and T. Hastie, "Regularization and variable selection via the elastic net," *Journal of the Royal Statistical Society Series B-Statistical Methodology,* vol. 67, pp. 301-320, 2005.

[24] J. Duchi and Y. Singer, "Efficient Online and Batch Learning Using Forward Backward Splitting," *Journal of Machine Learning Research,* vol. 10, pp. 2899-2934, Dec 2009.

[25] L. Xiao, "Dual Averaging Methods for Regularized Stochastic Learning and Online Optimization," *Journal of Machine Learning Research,* vol. 11, pp. 2543-2596, Oct 2010.

[26] J. Duchi, *et al.*, "Composite objective mirror descent," presented at the COLT, 2010.

[27] J. Langford, *et al.*, "Sparse Online Learning via Truncated Gradient," *Journal of Machine Learning Research,* vol. 10, pp. 777-801, Mar 2009.

[28] H. McMahan, "Follow-the-Regularized-Leader and Mirror Descent: Equivalence Theorems and L1 Regularization," presented at the AISTATS, 2011.

[29] I. Daubechies, *et al.*, "An iterative thresholding algorithm for linear inverse problems with a sparsity constraint," *Communications on Pure and Applied Mathematics,* vol. 57, pp. 1413-1457, Nov 2004.

[30] M. Elad, *et al.*, "A wide-angle view at iterated shrinkage algorithms," in *Proc. SPIE*, 2007, p. 670102.

[31] J. F. Cai, *et al.*, "Convergence of the Linearized Bregman Iteration for L(1)-Norm Minimization," *Mathematics of Computation,* vol. 78, pp. 2127-2136, Oct 2009.

[32] J. F. Cai, *et al.*, "Linearized Bregman Iterations for Compressed Sensing," *Mathematics of Computation,* vol. 78, pp. 1515-1536, Jul 2009.

[33] W. T. Yin, *et al.*, "Bregman Iterative Algorithms for l(1)-Minimization with Applications to Compressed Sensing," *Siam Journal on Imaging Sciences,* vol. 1, pp. 143-168, 2008.

[34] J. Pearl, "Embracing Causality in Default Reasoning," *Artificial Intelligence,* vol. 35, pp. 259-271, Jun 1988.

[35] P. Dayan and L. F. Abbott, *Theoretical Neuroscience. Computational and Mathematical Modeling of Neural System*. Cambridge, MA: MIT Press, 2001.

[36] C. Koch, *Biophysics of Computation*. New York: Oxford University Press, 1999.

[37] R. H. Masland, "The fundamental plan of the retina," *Nature Neuroscience,* vol. 4, pp. 877-86, Sep 2001.

[38] Available: http://media.xiph.org/video/derf/